\documentclass[5p]{elsarticle}

\journal{Computer Physics Communications}

\usepackage[english]{babel}
\usepackage[slantedGreek,sc]{mathpazo}
\usepackage{amsmath,amssymb,amsfonts}
\usepackage{graphicx}
\usepackage{booktabs}
\usepackage[pdftex,colorlinks=true,linkcolor=BrickRed,citecolor=ForestGreen,urlcolor=Blue]{hyperref}
\usepackage{color}
\usepackage{balance}

\def\eq#1{\begin{equation}#1\end{equation}}
\def\al#1{\begin{align}#1\end{align}}
\def\W{\operatorname{W}}
\def\O#1{\mathcal{O}(#1)}
\def\eps{\varepsilon}
\def\ie{e^{-1}}

\date{September 5, 2012}

\begin{document}

\begin{frontmatter}

\title{Lambert W Function for Applications in Physics}

\author[ung,ijs]{Darko Veberi\v{c}$~$}
\ead{darko.veberic@kit.edu}
\address[ung]{Laboratory for Astroparticle Physics, University of Nova
Gorica, Slovenia}
\address[ijs]{Department of Theoretical Physics, J.\ Stefan Institute,
Ljubljana, Slovenia}

\begin{abstract}
The Lambert $\W(x)$ function and its possible applications in physics 
are presented. The actual numerical implementation in C++ consists of 
Halley's and Fritsch's iterations with initial approximations based on 
branch-point expansion, asymptotic series, rational fits, and 
continued-logarithm recursion.
\end{abstract}


\end{frontmatter}

$~$
\\[1cm]
\begin{minipage}[t]{\textwidth}
\section*{Program summary}

\noindent
\textit{Program title:} LambertW\\
\textit{Catalogue identifier:} AENC\_v1\_0\\
\textit{Program summary URL:} \href{http://cpc.cs.qub.ac.uk/summaries/AENC_v1_0.html}{\tt http://cpc.cs.qub.ac.uk/summaries/AENC\_v1\_0.html}\\
\textit{Program obtainable from:} CPC Program Library, Queen's University, 
Belfast, N.~Ireland\\
\textit{Licensing provisions:} GNU General Public License version 3\\
\textit{No.\ of lines in distributed program, including test data, etc.:} 1335\\
\textit{No.\ of bytes in distributed program, including test data, etc.:} 25\,283\\
\textit{Distribution format:} \texttt{tar.gz}\\
\textit{Programming language:} C++ (with suitable wrappers it can be called 
from C, Fortran etc.), the supplied command-line\\
\phantom{012} utility is suitable for other scripting languages like sh, csh, 
awk, perl etc.\\
\textit{Computer:} All systems with a C++ compiler.\\
\textit{Operating system:} All Unix flavors, Windows. It might work with others.\\
\textit{RAM:} Small memory footprint, less than 1\,MB\\
\textit{Classification:} 1.1, 4.7, 11.3, 11.9.\\
\textit{Nature of problem:} Find fast and accurate numerical implementation for the Lambert W function.\\
\textit{Solution method:} Halley's and Fritsch's iterations with initial 
approximations based on branch-point expansion,\\
\phantom{012} asymptotic series, rational fits, and continued logarithm 
recursion.\\
\textit{Additional comments:} Distribution file contains the command-line 
utility \texttt{lambert-w}. Doxygen comments, included in\\
\phantom{012} the source files. Makefile.\\
\textit{Running time:} The tests provided take only a few seconds to run.\\
\textit{Source repository:} \href{https://github.com/DarkoVeberic/LambertW}{\tt https://github.com/DarkoVeberic/LambertW}
\end{minipage}

\clearpage

\section{Introduction}

The Lambert W function is defined as the inverse function of the
\eq{
x\mapsto x\,e^x
\label{map}
}
mapping and thus solves the
\eq{
y\,e^y = x
\label{def1}
}
equation. This solution is given in the form of the Lambert W function,
\eq{
y=\W(x),
\label{def2}
}
and according to Eq.~\eqref{def1} it satisfies the following relation,
\eq{
\W(x)\,e^{\W(x)}=x.
\label{definition}
}

\begin{figure}[t]
\centering
\includegraphics[width=\linewidth]{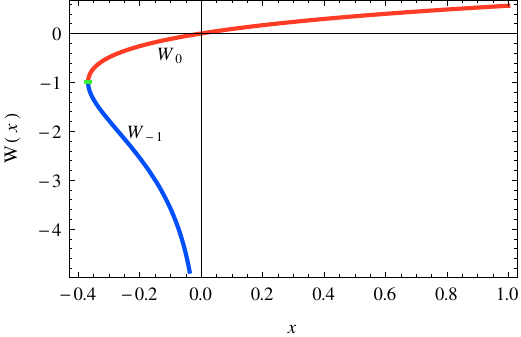}
\caption{The two branches of the Lambert W function, $\W_{-1}(x)$ in
blue and $\W_0(x)$ in red. The branching point at $(-\ie,\,-1)$ is
indicated with a green dash.}
\label{f:lambertW}
\end{figure}

Since the mapping in Eq.~\eqref{map} is not injective, no unique inverse 
of the $x\,e^x$ function exists (except at the minimum). As can be seen in 
Fig.~\ref{f:lambertW}, the Lambert W function has two real branches with a 
branching point located at $(-\ie,\,-1)$. The bottom branch, $\W_{-1}(x)$, is 
defined in the interval $x\in[-\ie,\,0]$ and has a negative singularity for 
$x\to0^-$.  The upper branch $\W_0(x)$ is defined for $x\in[-\ie,\,\infty]$.

The earliest mention of the problem of Eq.~\eqref{def1} is
attributed to Euler \cite{euler}. However, Euler himself credited 
Lambert for his previous work in this subject, Lambert's transcendental 
equation \cite{lambert}. The $\W(x)$ function started to be named after 
Lambert only recently, in the last 20 years or so; nevertheless, the 
first usage of the letter $\W$ appeared much earlier \cite{polya}.

Recently, the $\W(x)$ function amassed quite a following in the
mathematical community \cite{orcca}. Its most faithful proponents are 
suggesting elevating it among the present set of elementary functions, 
such as $\sin(x)$, $\cos(x)$, $\exp(x)$, $\ln(x)$, etc. The main 
argument for doing so is the fact that W is the root of the simplest 
exponential polynomial function given in Eq.~\eqref{def1}. We will 
acknowledge these efforts by strict usage of a \emph{roman} symbol W as 
its name.

While the Lambert W function is called \texttt{LambertW} in the 
mathematics software tool \emph{Maple} \cite{corless} and 
\texttt{lambertw} in the \textsc{Matlab} programming environment 
\cite{matlab}, in the \emph{Mathematica} computer algebra framework this 
function \cite{weisstein} is implemented under the name 
\texttt{ProductLog} (in the recent versions an alias \texttt{LambertW} 
is also supported). In open source format the Lambert W function is 
available in the special-functions part of the GNU Scientific Library 
(GSL) \cite{gsl}, unfortunately implemented using only the slower
Halley's iteration (see discussion in Section \ref{s:timing}).

There are numerous, well documented applications of $\W(x)$, certainly 
abundant in mathematics (like linear delay-differential equations 
\cite{asl}), numerics \cite{knuth}, computer science \cite{bustos} and 
engineering \cite{yi}, but surprisingly many also in physics, just to 
mention a few (without preference): quantum mechanics (solutions for 
double-well Dirac-delta potentials \cite{scott}), quantum statistics 
\cite{valluri}, general relativity (solutions to (1+1)-gravity problem 
\cite{scott2}, inverse of Eddington-Finkelstein (tortoise) coordinates 
\cite{eddington}), statistical mechanics \cite{caillol}, fluid dynamics 
\cite{pudasaini}, optics \cite{steinvall} etc.

The main motivation for the implementation in this work comes from the 
research in cosmic ray physics where it has been in use already for 
several years \cite{offline} and new interesting applications are 
appearing frequently \cite{unger}. In the next sections let us describe 
two new examples that arise from this field of astrophysics.

\subsection{Inverse of the Moyal function}

\begin{figure}[t]
\centering
\includegraphics[width=\linewidth]{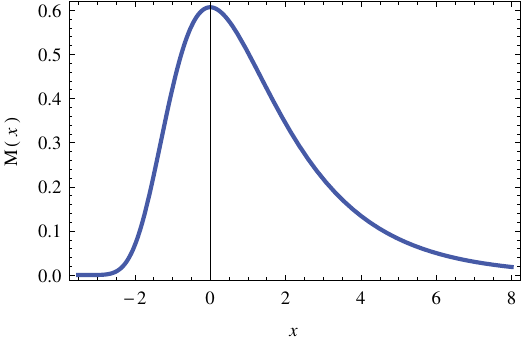}
\\
\includegraphics[width=\linewidth]{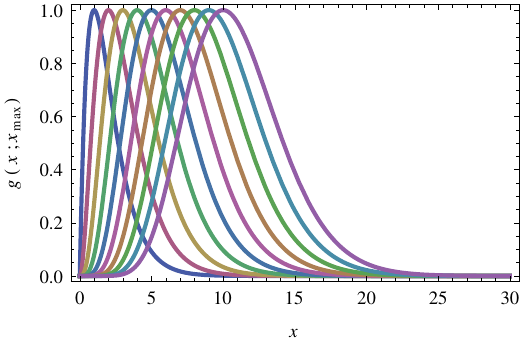}
\caption{\emph{Top:} The Moyal function $\operatorname{M}(x)$.
\emph{Bottom:} A family of one-parametric Gaisser-Hillas functions
$g(x;\,x_\text{max})$ for $x_\text{max}$ in the range from 1 to 10 with
step 1.}
\label{f:moyal-gh}
\end{figure}

The Moyal function and the related distribution was proposed as a good 
approximation for the more complicated Landau distribution \cite{moyal} 
used in descriptions of energy loss of charged particles passing through 
matter \cite{landau}. The un-normalized Moyal function is defined as
\eq{
\operatorname{M}(x)=\exp\left[-\tfrac12\left(x+e^{-x}\right)\right]
}
and can be seen in Fig.~\ref{f:moyal-gh} (top).

Its inverse can be written in terms of the two branches of the Lambert W
function,
\eq{
\operatorname{M}^{-1}_\pm(x)=
  \W_{0,-1}(-x^2) - 2\ln x.
}

In experimental astrophysics the Moyal function is used for 
phenomenological recovery of the saturated signals from photomultipliers 
\cite{maris}, where the largest parts of the saturated signals are 
obscured by the limited range of the analog-to-digital converters.

\subsection{Inverse of the Gaisser-Hillas function}

In astrophysics the Gaisser-Hillas function is used to model the 
longitudinal particle density in a cosmic-ray air shower \cite{gh}. We 
can show that the inverse of the three-parametric Gaisser-Hillas 
function,
\eq{
G(X;\,X_0,X_\text{max},\lambda) =
  \left[
    \frac{X-X_0}{X_\text{max}-X_0}
  \right]^{\frac{X_\text{max}-X_0}\lambda}
  \exp\left(\frac{X_\text{max}-X}\lambda\right),
}
is intimately related to the Lambert W function. Using rescale substitutions,
\eq{
x = \frac{X-X_0}\lambda\qquad\text{and}\qquad
x_\text{max} = \frac{X_\text{max}-X_0}\lambda,
}
the Gaisser-Hillas function is modified into a function of one parameter 
only,
\eq{
g(x;\,x_\text{max}) =
  \left[\frac{x}{x_\text{max}}\right]^{x_\text{max}}
  \exp(x_\text{max}-x).
}
The family of one-parametric Gaisser-Hillas functions is shown in 
Fig.~\ref{f:moyal-gh} (bottom). The problem of finding an inverse,
\eq{
g(x;\,x_\text{max}) \equiv y
}
for $0 < y \leqslant 1$, can be rewritten into
\eq{
-\frac{x}{x_\text{max}}\exp\left(-\frac{x}{x_\text{max}}\right) =
  -y^{1/x_\text{max}}\,\ie.
}
According to the definition \eqref{def1}, the two (real) solutions for 
$x$ are obtained from the two branches of the Lambert W function,
\eq{
x_{1,2} =
  -x_\text{max}\W_{0,-1}(-y^{1/x_\text{max}}\,\ie).
}
Note that the branch $-1$ or $0$ simply chooses the right or left side 
relative to the maximum, respectively.

The Gaisser-Hillas function is intimately related to the gamma distribution 
which was successfully used somewhat earlier \cite{longo} in an
approximate description of the mean longitudinal profile of the energy 
deposition in electromagnetic cascades. It is trivial to show that the 
inverses of the gamma and inverse-gamma distributions \cite{walck} are 
expressible in terms of the Lambert W function as well.

\section{Numerical methods}

Before describing the actual implementation let us review some of the 
possible numerical and analytical approaches to find $\W(x)$.

\subsection{Recursion}

For $x>0$ and $\W(x)>0$ we can take the natural logarithm of 
Eq.~\eqref{definition} and rearrange it into
\eq{
\W(x) = \ln x  - \ln\W(x).
\label{rear}
}
From here it is clear that an analytical expression for $\W(x)$
will exhibit a certain degree of self similarity. The $\W(x)$ function 
has multiple branches in the complex domain. Due to the $x>0$ and 
$\W(x)>0$ conditions, Eq.~\eqref{rear} represents the positive part 
of the $\W_0(x)$ principal branch and in this form it is suitable for 
evaluation when $\W_0(x)>1$, i.e.\ when $x>e$.

Unrolling the self-similarity in Eq.~\eqref{rear} as a recursive 
relation, the following curious expression for $\W_0(x)$ is obtained,
\eq{
\W_0(x) = \ln x - \ln(\ln x - \ln(\ln x - \,\ldots\,)),
\label{cle0}
}
or in the shorthand of a continued logarithm,
\eq{
\W_0(x) = \ln\frac{x}{\ln\frac{x}{\ln\frac{x}{\cdots}}}.
}
The above expression clearly has the form of successive approximations, 
the final result given by the limit, if it exists.

For $x<0$ and $\W(x)<0$ we can multiply both sides of 
Eq.~\eqref{definition} with $-1$, take logarithm, and rewrite it into a 
similar expansion for the $\W_{-1}(x)$ branch,
\eq{
\W(x) = \ln(-x) - \ln(-\W(x)).
}
Again, this leads to the similar recursion,
\eq{
\W_{-1}(x) = \ln(-x) - \ln(-(\ln(-x) - \ln(-\ldots))),
\label{clem1}
}
or as a continued logarithm,
\eq{
\W_{-1}(x) = \ln\frac{-x}{-\ln\frac{-x}{-\ln\frac{-x}{\cdots}}}.
\label{rec-w-1}
}
For this continued logarithm we will use the symbol $R_{-1}^{[n]}(x)$ 
where $n$ denotes the depth of the recursion in Eq.~\eqref{rec-w-1}.

Starting from yet another rearrangement of Eq.~\eqref{definition},
\eq{
\W(x)=\frac{x}{\exp\W(x)},
}
we can obtain a recursion relation for the $-\ie<x<e$ part of the 
principal branch $\W_0(x)<1$,
\eq{
\W_0(x) = \frac{x}{\exp\frac{x}{\exp\frac{x}{\ldots}}}.
}

\subsection{Halley's iteration}

We can apply Halley's root-finding method \cite{scavo} to derive an 
iteration scheme for $f(y)=W(y)-x$ by writing the second-order Taylor 
series
\eq{
f(y) = f(y_n) + f'(y_n)\,(y - y_n) + \tfrac12 f''(y_n)\,(y - y_n)^2
       + \cdots
\label{halley-second}
}
Since root $y$ of $f(y)$ satisfies $f(y)=0$ we can approximate the 
left-hand side of Eq.~\eqref{halley-second} with 0 and replace $y$ with 
$y_{n+1}$.  Rewriting the obtained result into
\eq{
y_{n+1} = y_n - \frac{f(y_n)}{f'(y_n)
          + \tfrac12 f''(y_n)\,(y_{n+1}-y_n)}
}
and using Newton's method $y_{n+1}-y_n=-f(y_n)/f''(y_n)$ on the last 
bracket, we arrive at the expression for Halley's iteration for the
Lambert W function
\eq{
W_{n+1} = W_n + \frac{t_n}{t_n\,s_n - u_n},
}
where
\al{
t_n &= W_n\,e^{W_n} - x,
\\
s_n &= \frac{W_n+2}{2(W_n+1)},
\\
u_n &= (W_n+1)\,e^{W_n}.
}

This method is of the third order, i.e.\ having $W_n=\W(x)+\O\eps$ will 
produce $W_{n+1}=\W(x)+\O{\eps^3}$. Supplying this iteration with a
sufficiently accurate first approximation of the order of $\O{10^{-4}}$ 
will thus give a machine-size floating point precision $\O{10^{-16}}$ in 
at least two iterations.

\subsection{Fritsch's iteration}

For both branches of the Lambert W function a more efficient iteration scheme 
exists \cite{fritsch},
\eq{
W_{n+1}=W_n(1+\eps_n),
}
where $\eps_n$ is the relative difference of successive approximations at 
iteration $n$,
\eq{
\eps_n=\frac{W_{n+1}-W_n}{W_n}.
}
The relative difference can be expressed as
\eq{
\eps_n = 
  \left(\frac{z_n}{1+W_n}\right)
  \left(\frac{q_n-z_n}{q_n-2z_n}\right),
}
where
\al{
z_n&=\ln\frac{x}{W_n}-W_n,
\\
q_n&=2(1+W_n)\left(1+W_n+\tfrac23z_n\right).
}
The error term in this iteration is of a fourth order, i.e. with 
$W_n=\W(x)+\O{\eps_n}$ we obtain $W_{n+1}=\W(x)+\O{\eps_n^4}$.

Supplying this iteration with a sufficiently reasonable first guess,
accurate to the order of $\O{10^{-4}}$, will therefore deliver
machine-size floating point precision $\O{10^{-16}}$ in only one 
iteration and excessive $\O{10^{-64}}$ in two! The main goal now is to find 
reliable first order approximations that can be fed into Fritsch's 
iteration. Due to the lively landscape of the Lambert W function properties, 
the approximations will have to be found in all the particular ranges of the 
function's behavior.

\section{Initial approximations}

The following section deals with finding the appropriate initial
approximations in the whole definition range of the two branches of the
Lambert W function.

\subsection{Branch-point expansion}

\begin{figure}[t]
\centering
\includegraphics[width=\linewidth]{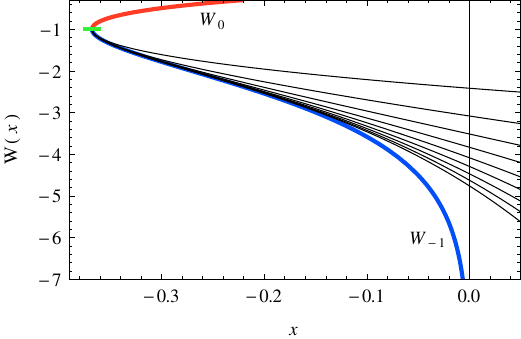}
\\
\includegraphics[width=\linewidth]{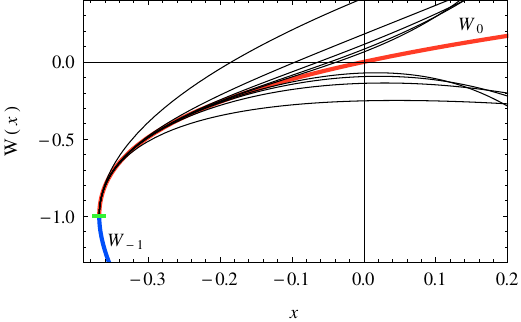}
\caption{Successive orders of the branch-point expansion for
$\W_{-1}(x)$ on the top and $\W_0(x)$ on the bottom.}
\label{f:bp-exp}
\end{figure}

The inverse of the Lambert W function, $\W^{-1}(y)=y\,e^y$, has
two extrema located at $\W^{-1}(-1)=-\ie$ and $\W^{-1}(-\infty)=0^-$.  
Expanding $\W^{-1}(y)$ to the second order around the minimum at $y=-1$ 
we obtain
\eq{
\W^{-1}(y) \approx -\frac{1}{e}+\frac{(y+1)^2}{2e}.
}
The inverse $\W^{-1}(y)$ is thus approximated in the lowest order by a 
parabolic term implying that the Lambert W function will have 
square-root behavior in the vicinity of the branch point $x=-\ie$,
\eq{
\W_{-1,0}(x) \approx -1 \mp \sqrt{2(1+ex)}.
\label{bp-first}
}
To obtain the additional terms in Eq.~\eqref{bp-first} we proceed by 
defining an inverse function, centered and rescaled around the minimum, 
i.e.\ $f(y)=2(e\W^{-1}(y-1)+1)$. Due to this centering and rescaling, 
the Taylor series around $y=0$ becomes particularly neat,
\eq{
f(y) = y^2 + \tfrac23y^3 + \tfrac14y^4 + \tfrac1{15}y^5 + \cdots
}
Using this expansion we can derive coefficients \cite{morse} of the 
corresponding inverse function
\al{
f^{-1}(z) &= 1 + \W\left(\frac{z-2}{2e}\right) =
\label{bp-inverse}
\\
  &= z^{1/2} - \tfrac13z + \tfrac{11}{72}z^{3/2} - \tfrac{43}{540}z^2 + 
     \cdots
}
From Eq.~\eqref{bp-inverse} we see that $z=2(1+ex)$. Using 
$p_\pm(x)=\pm\sqrt{2(1+ex)}$ as an independent variable we can write this 
series expansion as
\eq{
\W_{-1,0}(x) \approx
  B_{-1,0}^{[n]}(x) =
    \sum_{i=0}^n b_i p_\mp^i(x),
\label{bp}
}
where the lowest few coefficients $b_i$ are
\begin{center}
\begin{tabular}{lrrrrrrrr}
\toprule
$i$ & 0 & 1 & 2 & 3 & 4 & 5 & 6 & 7
\\
\midrule
$b_i$ & -1 & 1 & $-\tfrac13$ & $\tfrac{11}{72}$ & $-\tfrac{43}{540}$ &
  $\tfrac{769}{17\,280}$ & $-\tfrac{221}{8\,505}$ & 
  $\tfrac{680\,863}{43\,545\,600}$
\\
\bottomrule
\end{tabular}
\end{center}
and more of them are given in the accompanying code (see Fig.~\ref{f:bp-exp}
for succesive orders of the series).

\subsection{Asymptotic series}

Another useful tool is the asymptotic expansion \cite{debruijn}. Using
\eq{
A(a,b) = a - b + \sum_k \sum_m C_{km}a^{-k-m-1}b^{m+1},
}
with $C_{km}$ related to the Stirling numbers of the first kind, the 
Lambert W function can be expressed as
\eq{
\W_{-1,0}(x) = A(\ln(\mp x),\,\ln(\mp\ln(\mp x))),
}
with $a=\ln x$, $b=\ln\ln x$ for the $\W_0$ branch and $a=\ln(-x)$, 
$b=\ln(-\ln(-x))$ for the $\W_{-1}$ branch. The first few terms are
\al{
A(a,b) & =
  a - b + \frac{b}{a} + \frac{b(-2+b)}{2a^2} +
  \frac{b(6-9b+2b^2)}{6a^3} +
\nonumber
\\
  & \, +
  \frac{b(-12+36b-22b^2+3b^3)}{12a^4} +
\label{asym}
\\
  & \, +
  \frac{b(60-300b+350b^2-125b^3+12b^4)}{60a^5} + \cdots
\nonumber
}

\subsection{Rational fits}

A useful quick-and-dirty approach to the functional approximation is to 
generate a large enough sample of data points $\{w_i\,e^{w_i},\,w_i\}$, which
evidently lie on the Lambert W function. Within some appropriately chosen range 
of $w_i$ values, the points are fitted with a rational approximation
\eq{
Q(x)=\frac{\sum_i a_i x^i}{\sum_i b_i x^i},
}
varying the order of the polynomials in the nominator and denominator, 
and choosing the one that has the lowest maximal absolute residual in a 
particular interval of interest.

For the $\W_0(x)$ branch, the first set of equally-spaced $w_i$ 
components was chosen in a range that produced $w_i\,e^{w_i}$ values in 
an interval $[-0.3,\,0]$. The optimal rational fit turned out to be
\eq{
Q_0(x) =
 x \frac{1+a_1x+a_2x^2+a_3x^3+a_4x^4}
        {1+b_1x+b_2x^2+b_3x^3+b_4x^4}
\label{rat}
}
where the coefficients\footnote{The ' symbol in coefficient values 
denotes truncation in this presentation; the full machine-size sets of 
decimal places are given in the accompanying code.} for this first 
approximation $Q_0^{[1]}(x)$ are
\begin{center}
\begin{tabular}{lrrrr}
\toprule
$i$ & 1 & 2 & 3 & 4
\\
\midrule
$a_i$ &
5.931375' & 11.392205' & \phantom{0}7.338883' & 0.653449'
\\
$b_i$ &
6.931373' & 16.823494' & 16.430723' & 5.115235'
\\
\bottomrule
\end{tabular}
\end{center}

For the second fit of the $\W_0(x)$ branch a $w_i$ range was chosen so 
that the $w_i\,e^{w_i}$ values were produced in the interval 
$[0.3,\,2e]$, giving rise to the second optimal rational fit 
$Q_0^{[2]}(x)$ of the same form as in Eq.~\eqref{rat} but with 
coefficients
\begin{center}
\begin{tabular}{lrrrr}
\toprule
$i$ & 1 & 2 & 3 & 4
\\
\midrule
$a_i$ &
2.445053' & 1.343664' & 0.148440' & 0.000804'
\\
$b_i$ &
3.444708' & 3.292489' & 0.916460' & 0.053068'
\\
\bottomrule
\end{tabular}
\end{center}

For the $\W_{-1}(x)$ branch a single rational approximation of the form
\eq{
Q_{-1}(x) =
  \frac{a_0+a_1x+a_2x^2}
       {1+b_1x+b_2x^2+b_3x^3+b_4x^4+b_5x^5}
\label{rat-1}
}
with the coefficients
\begin{center}
\begin{tabular}{lrrr}
\toprule
$i$ & 0 & 1 & 2
\\
\midrule
$a_i$ & -7.814176' & 253.888101' & 657.949317'
\\
$b_i$ & & -60.439587' & 99.985670'
\\
\toprule
$i$ & 3 & 4 & 5
\\
\midrule
$b_i$ & 682.607399' & 962.178439' & 1477.934128'
\\
\bottomrule
\end{tabular}
\end{center}
is enough.

\section{Implementation}

To quantify the accuracy of a particular approximation $\widetilde\W(x)$ 
of the Lambert function $\W(x)$ we can introduce a quantity $\Delta(x)$ 
defined as
\eq{
\Delta(x)=
  \log_{10}|\W(x)| -
  \log_{10}|\widetilde\W(x)-\W(x)|,
}
so that it gives us a number of correct decimal places the approximation 
$\widetilde\W(x)$ is producing for some value of the parameter $x$.

\begin{figure}[t]
\centering
\includegraphics[width=\linewidth]{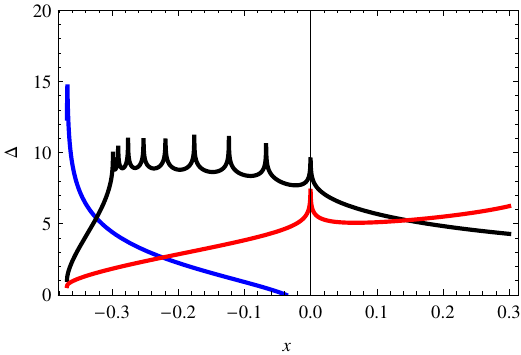}
\\
\includegraphics[width=\linewidth]{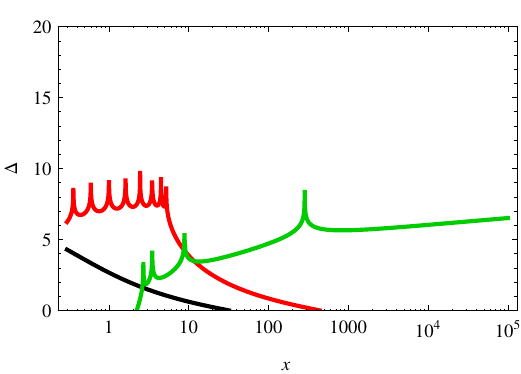}
\caption{Combining different approximations of $\W_0(x)$ into a final
piecewise function. The number of accurate decimal places $\Delta(x)$ is
shown for two ranges, linear interval $[-\ie,\,0.3]$ (top) and
logarithmic interval $[0.3,\,10^5]$ (bottom). The approximations are
branch-point expansion $B_0^{[9]}(x)$ from Eq.~\eqref{bp} (blue),
rational fits $Q_0^{[1]}(x)$ and $Q_0^{[2]}(x)$ from Eq.~\eqref{rat} in
black and red, respectively, and asymptotic series $A_0(x)$ from
Eq.~\eqref{asym} (green).}
\label{f:approx-w0}
\end{figure}

In Fig.~\ref{f:approx-w0} all approximations for the $\W_0(x)$ mentioned 
above are shown in the linear interval $[-\ie,\,0.3]$ on the left and the
logarithmic interval $[0.3,\,10^5]$ on the right. For each of the 
approximations an optimal interval is chosen so that the number of 
accurate decimal places is maximized over the whole definition range.  
For the $\W_0(x)$ branch the resulting piecewise approximation
\eq{
\widetilde\W_0(x)=
\begin{cases}
B_0^{[9]}(x) & ;\, -\ie \leqslant x < a
\\
Q_0^{[1]}(x) & ;\, a \leqslant x < b
\\
Q_0^{[2]}(x) & ;\, b \leqslant x < c
\\
A_0(x) & ;\,c \leqslant x < \infty
\end{cases}
\label{approx-w0}
}
with $a=-0.323581'$, $b=0.145469'$, and $c=8.706658'$, is accurate in 
the definition range $[-\ie,\,7]$ to at least 5 decimal places and to at 
least 3 decimal places in the whole definition range $[-\ie,\,\infty]$.  
Note that $B_0^{[9]}(x)$ comes from Eq.~\eqref{bp}, $Q_0^{[1]}(x)$ 
and $Q_0^{[2]}(x)$ are from Eq.~\eqref{rat}, and $A_0(x)$ is from 
Eq.~\eqref{asym}.

\begin{figure}[t]
\centering
\includegraphics[width=\linewidth]{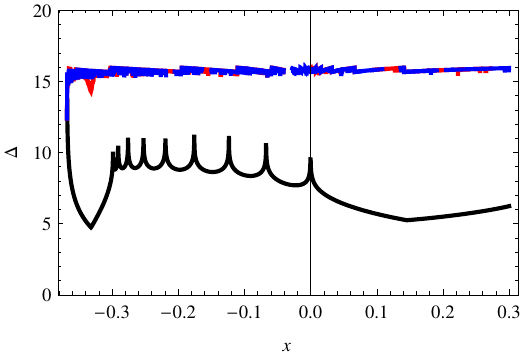}
\\
\includegraphics[width=\linewidth]{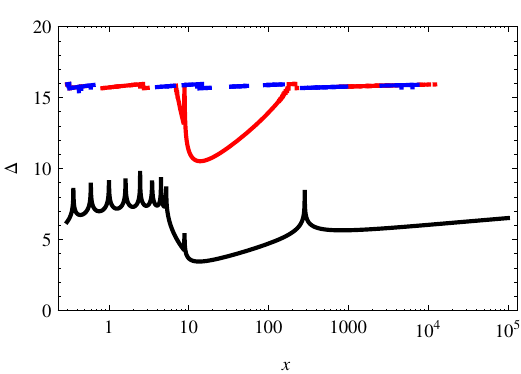}
\caption{Final values of the combined approximation $\widetilde\W_0(x)$
(black) from Fig.~\ref{f:approx-w0} after one step of Halley's
iteration (red) and one step of Fritsch's iteration (blue).}
\label{f:approx-w0-step}
\end{figure}

The accuracy of the final piecewise approximation $\widetilde\W_0(x)$ is 
shown in Fig.~\ref{f:approx-w0-step} (black line). Using this approximation, a 
single step of Halley's iteration (red) and Fritsch's iteration (blue) 
is performed and the resulting number of accurate decimal places is shown. As 
can be seen from the plots, both iterations produce machine-size accurate 
floating point numbers in the whole definition interval except for the interval 
between 6.5 and 190 where the Halley's method requires another step of the 
iteration. For that reason we have decided to use only (one step of) 
Fritsch's iteration in the C++ implementation of the Lambert W function.

\begin{figure}[t]
\centering
\includegraphics[width=\linewidth]{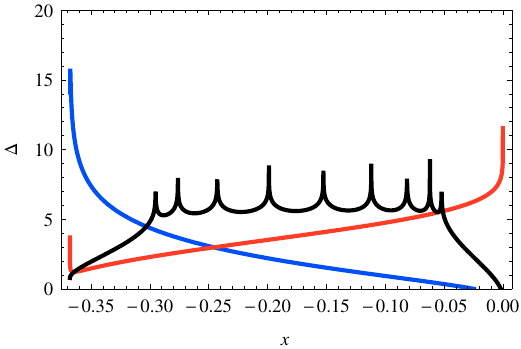}
\\
\includegraphics[width=\linewidth]{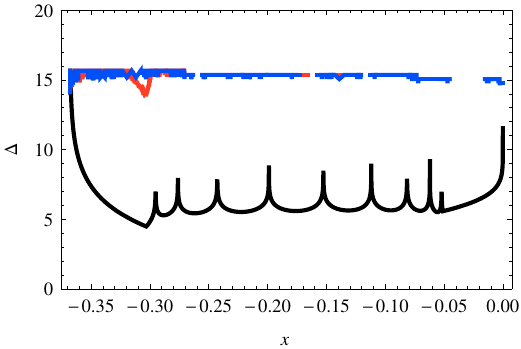}
\caption{\emph{Top:} Approximations of the $\W_{-1}(x)$ branch. The
branch-point expansion $B_{-1}^{[9]}(x)$ is shown in blue, the rational
approximation $Q_{-1}(x)$ in black, and the logarithmic recursion
$R_{-1}^{[9]}$ in red. \emph{Bottom:} The combined approximation is accurate
to at least 5 decimal places in the whole definition range. The results
after applying one step of Halley's iteration are shown in red and
after one step of Fritsch's iteration in blue.}
\label{f:approx-w-1}
\end{figure}

In Fig.~\ref{f:approx-w-1} (top) the same procedure is shown for the 
$\W_{-1}(x)$ branch. The final approximation
\eq{
\widetilde\W_{-1}(x)=
\begin{cases}
B_{-1}^{[9]}(x) & ;\, -\ie \leqslant x < a
\\
Q_{-1}(x) & ;\, a \leqslant x < b
\\
R_{-1}^{[9]}(x) & ;\, b \leqslant x < 0
\end{cases}
\label{approxw-1}
}
with $a=-0.302985'$ and $b=-0.051012'$, is accurate to at least 5 
decimal places in the whole definition range $[-\ie,\,0]$. Note that 
$B_{-1}^{[9]}(x)$ is taken from Eq.~\eqref{bp}, $Q_{-1}(x)$ is from 
Eq.~\eqref{rat-1}, and $R_{-1}^{[9]}(x)$ is from Eq.~\eqref{rec-w-1}.

In Fig.~\ref{f:approx-w-1} (bottom) the combined approximation 
$\widetilde\W_{-1}(x)$ is shown (black line). The values after one step of
Halley's iteration are shown in red and after one step of Fritsch's 
iteration in blue. Similarly as for the previous branch, Fritsch's 
iteration turns out to be superior, yielding machine-size accurate results in 
the whole definition range, while Halley's iteration is accurate to at 
least 13 decimal places.

\section{Source availability, installation and usage}

The most recent version of the sources of this implementation with some 
additional material and examples are available from
\\
\href{https://github.com/DarkoVeberic/LambertW}{\tt https://github.com/DarkoVeberic/LambertW}
\\
and are released under the dual GPL/BSD license.

Apart from the special functions in GSL \cite{gsl}, this is the only freely 
available implementation of the Lambert W function in C++ and to the best of 
our knowledge the only implementation using the superior Fritsch's version of 
the iteration.

The supplied C++ code implements the following set of functions\footnote{Which 
can be found in the files \texttt{LambertW.h} and \texttt{LambertW.cc}.}
\begin{itemize}
\item \texttt{double LambertWApproximation<$b$>(double x);}
\item \texttt{double LambertW<$b$>(double x);}
\item \texttt{double LambertW(int branch, double x);}
\end{itemize}
where $b$ in the first two functions should be replaced with the 
compile-time branch integer value, e.g.\ \texttt{LambertW<-1>(x)} or 
\texttt{LambertW<0>(x)}. Apart from the slightly increased efficiency, 
the main reason for implementing the first two functions with the branch $b$ as 
a compile time parameter is to force the users to think about the two possible 
solutions to the problem in Eq.~\eqref{def1}.  Just as for solutions to the 
quadratic equation where the $\pm$ sign has to be chosen based on the desired 
solution, the Lambert W function offers two possibilities that need careful 
consideration.

The initial approximations $\widetilde\W_b(x)$ from Eqs.~\eqref{approxw-1} and 
\eqref{approx-w0}, that are used to kick-start the iterations, are also 
directly available as \texttt{LambertWApproximation<$b$>(x)}, as they might be 
useful in applications for which it is sufficient to have a limited number of
accurate decimal places (see the discussion above).

The supplied code does not need any kind of special installation procedures. In 
the case that your analysis needs an evaluation of the Lambert W function, the 
two source files, \texttt{LambertW.h} and \texttt{LambertW.cc}, should be 
simply bundled with your project structure and compiled with a suitable C++ 
compiler.

The source distribution also includes a command-line utility implemented by the 
\texttt{lambert-w.cc} source file. The included \texttt{Makefile} can, with the 
request \texttt{make lambert-w}, build an executable command. It can be invoked 
through a shell as \texttt{./lambert-w [branch] x}, taking an optional 
\texttt{branch} number (0 by default) and a parameter \texttt{x}. The output of 
the command is equivalent to the $\W_\texttt{branch}(\texttt{x})$ return value 
and can thus be simply used in shell scripts (\texttt{sh}, \texttt{bash}, or 
\texttt{csh}) or other programming languages with easy access to shell 
invocations (\texttt{awk}, \texttt{perl} etc.).

Also included in the distribution is a test suite which can perform a
correctness check on your build by comparing obtained and expected return 
values of the Lambert W function on your system. It is invoked by the command 
\texttt{make tests}. Any potential discrepancies larger than the double machine 
precision ($\gtrsim10^{-14}$) will be reported in the output.

\section{Timing}
\label{s:timing}

\begin{figure}[t]
\centering
\includegraphics[width=\linewidth]{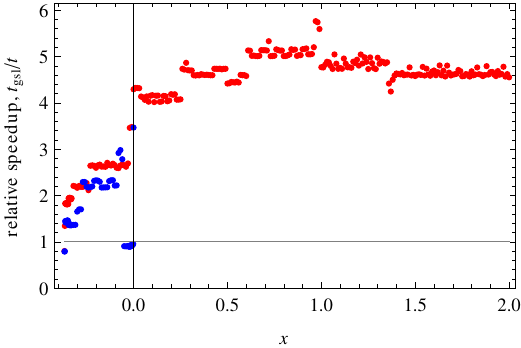}
\caption{Execution speedup $t'_\text{gsl}/t'$, relative to the GSL
implementation \cite{gsl} for the $\W_0(x)$ branch (red) and the $\W_{-1}(x)$
branch (blue). For $x>2$ the ratio slowly decreases and is $\sim2$ for $x>8$.
Time of the respective driver loops and function calls was subtracted in order
to measure only differences between the pure numerical parts of the two
implementations (see text for details).}
\label{f:timing}
\end{figure}

We have decided to compare the execution time of our code relative to the GNU 
GSL library (implemented in the C language) since comparisons to interpreted
code (like \textit{Maple}, \textsc{Matlab} or \textit{Mathematica}) would,
besides common availability problems, not be fair in terms of speed.

In Fig.~\ref{f:timing} relative speedup, $t_\text{gsl}/t$, is shown as a 
function of the Lambert W parameter $x$. Timing accuracy of several \% was 
achieved by running 3\,000\,000 function calls in a loop, taking special care 
that the compiler did not optimize code away by slightly modifying $x$ on each 
call and then gathering all results into a summed variable reported at the end.  
For each of the two implementations an identical code was also run with the 
Lambert W function call replaced with a simple identity function (just 
returning its input parameter) in order to estimate the overhead of the 
surrounding timing code.  This $t_\text{overhead}$ is then consequently 
subtracted from the time of the Lambert W runs $t$, giving an approximation to 
the time taken by the pure function call, $t'=t-t_\text{overhead}$. The ratio 
$t'_\text{gsl}/t'$ is then plotted in Fig.~\ref{f:timing}.

As can be clearly seen from Fig.~\ref{f:timing}, our implementation is at least 
$2\times$ faster than GSL for a broad range of input parameters $x$, but the 
largest efficiency gains (up to $5\times$) are observed in the ranges where 
rational fits $Q_b$ from Eqs.~\eqref{approx-w0} and \eqref{approxw-1} are used.  
Although Fritsch's iteration is in general more complex than Halley's, at the 
end it pays off, yielding machine-size accuracy with a single step where 
Halley's might need more, also due to poor initial approximations used in GSL.  
GSL performs better only in the small regions where branch-point and asymptotic 
expansions are used without the consequent Halley's iteration refinements.  The 
comparisons were made on the Ubuntu 12.04 x86\_64 Linux operating system
running on a 2.2\,GHz AMD Opteron 275 processor, using the GCC 4.6.3 compiler
with optimization option \texttt{-O2}.

\section{Conclusions}

We have shown that Fritsch's iteration scheme coupled with accurate initial 
approximations can deliver significant efficiency gains in the numerical 
evaluation of the real branches of the Lambert W function. The open-source 
release of our C++ implementation is suitable for inclusion in various analysis 
packages used in all fields of physics.

\section*{Acknowledgments}

The author wishes to thank Matej Horvat, Michael Unger, Martin O'Loughlin,
and Amir Malekpour for useful discussions and suggestions. This work was 
partially supported by the Slovenian Research Agency.

\end{document}